\documentclass[sigconf,nonacm, screen]{acmart}

\usepackage{stfloats}
\usepackage{xcolor}
\usepackage{enumitem}

\AtBeginDocument{%
  \providecommand\BibTeX{{%
    \normalfont B\kern-0.5em{\scshape i\kern-0.25em b}\kern-0.8em\TeX}}}

\setcopyright{acmlicensed}
\copyrightyear{2018}
\acmYear{2018}
\acmDOI{XXXXXXX.XXXXXXX}

\acmConference[Conference acronym 'XX]{Make sure to enter the correct
  conference title from your rights confirmation emai}{June 03--05,
  2018}{Woodstock, NY}
\acmBooktitle{Woodstock '18: ACM Symposium on Neural Gaze Detection,
 June 03--05, 2018, Woodstock, NY} 
\acmISBN{978-1-4503-XXXX-X/18/06}

\definecolor{mypink}{RGB}{241, 225, 225}
\definecolor{myblue}{RGB}{204, 228, 240}

\newcommand{\method}[1]{Language-Oriented Code Sketching}

\newsavebox{\coloredquotationbox}
\newenvironment{coloredquotation}
 {%
  \begin{trivlist}
  \item\relax
  \begin{lrbox}{\coloredquotationbox}
  \setlength{\fboxsep}{10pt} 
  \begin{minipage}{\dimexpr\linewidth-2\fboxsep}
 }
 {%
  \end{minipage}
  \end{lrbox}
  \item\relax
  \parbox{\linewidth}{
    \begingroup
    \color[RGB]{224,215,188}%
    \hrule
    \color[RGB]{249,245,233}%
    \hrule
    \color[RGB]{224,215,188}%
    \hrule
    \endgroup
    
    \setlength{\fboxsep}{10pt} 
    \colorbox[RGB]{249,245,233}{\usebox{\coloredquotationbox}}\par\nointerlineskip
    
    \begingroup
    \color[RGB]{224,215,188}%
    \hrule
    \color[RGB]{249,245,233}%
    \hrule
    \color[RGB]{224,215,188}%
    \hrule
    \endgroup
  }
  \end{trivlist}
 }

\usepackage{textcomp}
\begin{document}

\settopmatter{authorsperrow=4}

\title{Sketch Then Generate: Providing Incremental User Feedback and Guiding LLM Code Generation through \\ Language-Oriented Code Sketches}

\author{Chen Zhu-Tian}
\email{ztchen@umn.edu}
\orcid{0000-0002-2313-0612}
\affiliation{%
  \institution{University of Minnesota}
  \city{Minneapolis}
  \state{MN}
  \country{USA}
}

\author{Zeyu Xiong}
\email{zeyuxionghci@outlook.com}
\orcid{0000-0002-3652-1890}
\affiliation{%
  \institution{University of Minnesota}
  \city{Minneapolis}
  \state{MN}
  \country{USA}
}

\author{Xiaoshuo Yao}
\email{yaoxiaos@usc.edu}
 \orcid{0000-0002-9740-7506}
\affiliation{%
  \institution{USC}
  \city{Los Angeles}
  \state{CA}
  \country{USA}
}

\author{Elena Glassman}
\email{glassman@seas.harvard.edu}
\orcid{0000-0001-5178-3496}
\affiliation{%
  \institution{Harvard University}
  \city{Boston}
  \state{MA}
  \country{USA}
}
\begin{abstract}
Crafting effective prompts for code generation or editing with Large Language Models (LLMs) is not an easy task.
Particularly, the absence of immediate, stable feedback during prompt crafting hinders effective interaction, as users are left to mentally imagine possible outcomes until the code is generated. In response, we introduce \emph{Language-Oriented Code Sketching}, an interactive approach that provides instant, incremental feedback in the form of code sketches (i.e., incomplete code outlines) during prompt crafting. 
This approach converts a prompt into a code sketch by 
leveraging the inherent linguistic structures within the prompt 
and applying classic natural language processing techniques.
The sketch then serves as an intermediate placeholder that 
not only previews the intended code structure but also guides the LLM towards the desired code, 
thereby enhancing human-LLM interaction. 
We conclude by discussing the approach's applicability and future plans.
\end{abstract}


\begin{teaserfigure}
    \hspace{-4mm}
  \includegraphics[width=1.04\textwidth]{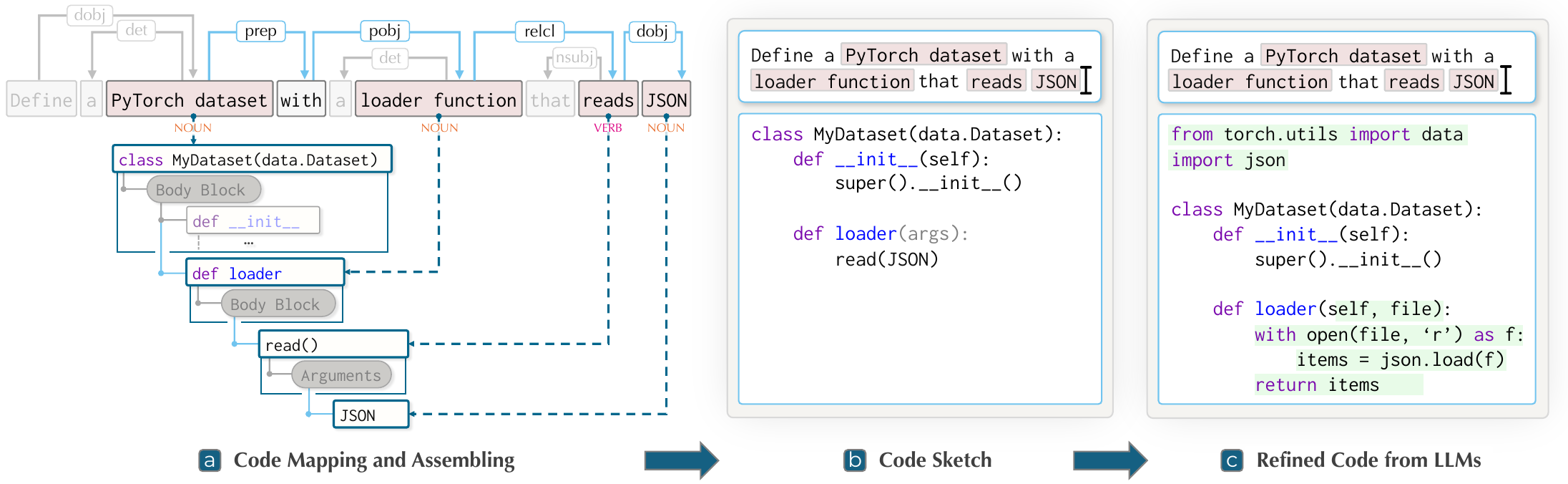}
  \caption{We derive a code sketch from the user's prompt a) by mapping the phrases to code elements then assembling them.
    The code sketch can b) offer user feedback on the intended code structure and c) guide LLMs in generating the final code.    
    }
  \Description{Enjoying the baseball game from the third-base
  seats. Ichiro Suzuki preparing to bat.}
  \label{fig:teaser}
\end{teaserfigure}


\maketitle

\section{Introduction}

The advent of Large Language Models (LLMs) represents a paradigm shift in programming, 
enabling an era where natural language prompts can directly inform code generation and editing~\cite{zan2023large, zhang2024unifying}. 
Despite this advancement,
programmers frequently encounter a tedious loop of trial and error, continuously refining their prompts to guide LLMs to the desired code, e.g., as shown in the user studies of GitHub Copilot~\cite{vaithilingam2022expectation, barke2023grounded, liang2023largescale}.
Existing solutions (e.g.,~\cite{yen2023coladder, liu2023wants, ross2023programmer, tian2023interactive}), 
while useful, 
fall short in providing user feedback during the prompt crafting process. 
The lack of such feedback can lead to uncertainty about the expected output and potential inefficiencies in the development process, 
as users are left to mentally imagine possible outcomes until the code is generated~\cite{10.1145/3544548.3580817, 10.1145/3487569}.

However, providing user feedback during prompt crafting presents significant challenges.
Ideally, feedback should not only be \emph{instant} but also evolve \emph{incrementally} as the user types their prompt, 
offering clear and understandable intermediate results that progressively guide the LLM towards the final desired code.
The most straightforward approach—continuously sending the user's prompt to LLMs for feedback—is currently impractical, due to the computational demands and the unpredictable nature of LLM outputs, 
which can vary greatly as the prompt evolves~\cite{vaithilingam2022expectation}.
These limitations highlight the need for innovative approaches that extend beyond mere reliance on LLMs.

One observation, however, is that users' prompts often include phrases directly referencing code elements.
Moreover, these phrases are embedded within narratives that reflect the relationships among these elements in the intended code structure.
For example, consider the prompt \emph{``Define a PyTorch dataset with a loader function ...''} (\autoref{fig:teaser}a).
Here, \colorbox{mypink}{\texttt{PyTorch dataset}} and \colorbox{mypink}{\texttt{loader function}} reference a class and a function, respectively.
Their linguistic relationship --- where \colorbox{mypink}{\texttt{loader function}} is a child of \colorbox{mypink}{\texttt{PyTorch dataset}} associated through \emph{``with''} --- reveals that the function is meant to be a component of the class.
This linguistic structure of the prompt offers an interesting opportunity to derive a \emph{code sketch} --- an incomplete code outline~\cite{solar2013program}) --- by mapping these phrases to predefined code elements and assembling them based on their linguistic relationship.
The resultant sketch (\autoref{fig:teaser}b) not only offers a preview of the intended code structure of the prompt,
but also serves as a foundational layer for LLMs to further refine and complete (\autoref{fig:teaser}c).
Such a process can be achieved using classic natural language processing (NLP) techniques, 
bypassing the constraints inherent in solutions based solely on LLMs.

Based on the observation,
we develop \textbf{Language-Oriented Code Sketching}, 
an interactive approach that incrementally derives a code sketch as the user types a prompt.
We next introduce the technical details of our approach and discuss its applicability, along with user feedback collected through a technology probe.

\section{Related Work}

The interaction between programmer and LLMs in coding
has been the subject of extensive research, e.g.,~\cite{Kazemitabaar2023,nguyen2024beginning,prather2023, barke2023grounded, sarkar2022like, vaithilingam2022expectation, liang2023largescale, mozannar2022reading}. 
Among the challenges in coding with AI assistants,
the crafting of effective prompts has repeatedly been identified as a primary concern. 
This challenge stems from the need for a deep understanding of LLMs' interpretative mechanisms and the ability to articulate intentions in natural language. 
To address this challenge, researchers have develop various interactive systems and approaches to 
assist users in writing effective prompts~\cite{yen2023coladder, angert2023spellburst, ross2023programmer, cai2023low, liu2023wants, tian2023interactive}.
For instance, Coladder~\cite{yen2023coladder} introduces a block-based user interface that enables programmers to hierarchically decompose and refine their prompts.
Liu el al.~\cite{liu2023wants} proposed Grounded Code Matching, a method where the LLM generates code and accompanying explanations, allowing users to refine these explanations to iteratively improve the LLM's output. 
Despite useful, these solutions fall short in providing user feedback during the prompt crafting process.
The slow feedback loop between writing a prompt, submitting it, and awaiting the generated code can undermines productivity and escalates frustration. 

This work aims to provide instant, incremental user feedback during the prompt crafting process
by deriving code sketches from the latent linguistic structures of the prompts on-the-fly.
Our approach shares a similar spirit to methods in semantic parsing~\cite{kamath2018survey}, which translates natural language into formal, executable structures, e.g., SQL or code. 
However, our approach is different in two key aspects: 1) we focus on generating incomplete code sketches for user feedback rather than executable code; and 2) our method emerges naturally in the user's prompt typing process, without imposing additional effort on the user.
In this sense, 
our method is closer to systems that treats the natural language as implicit data sources
to foster fluid and streamlined interactions. 
Such a method has been applied in various domains, such as creating data reports~\cite{DBLP:conf/chi/ChenX22}, unit animations~\cite{DBLP:conf/chi/CaoECX23}, augmented videos~\cite{DBLP:journals/tvcg/ChenYXBXWP23}, and graphics~\cite{DBLP:conf/uist/Xia20}.
Additionally,
inspired by structural code editors~\cite{donzeau1980programming}, 
when deriving the sketch,
we manipulate the underlying abstract syntax tree of the code sketch, instead of directly editing its textual representation.
This enables us to fully leverage the structures in both the prompt and the code itself.

\section{\method{}}

\begin{figure}[h]
    \centering
    \vspace{-4mm}
    \includegraphics[width=1\columnwidth]{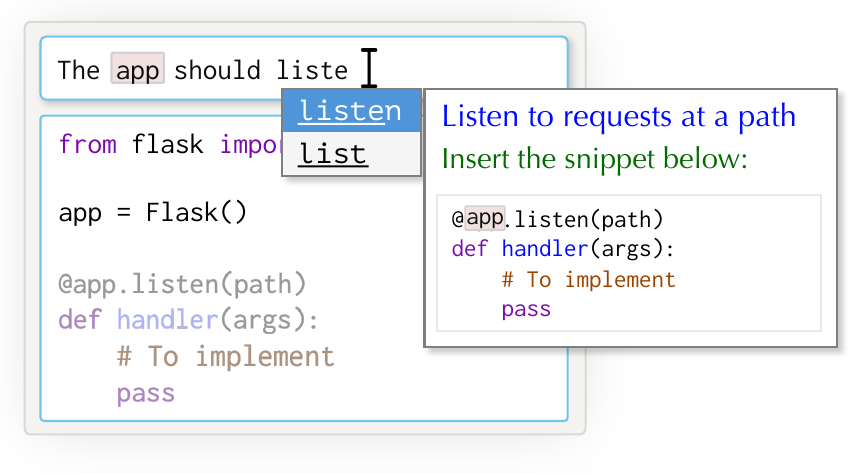}
    \vspace{-6mm}
    \caption{
    As the user types a phrase, our system suggests corresponding code elements and previews their integration into the existing code. When the user accepts a suggestion, the code element is inserted and the association between the phrase and the code element is preserved.
    }
    \label{fig:ui_exp}
\end{figure}


\noindent
Our approach leverages word completion as an interface (\autoref{fig:ui_exp})
to facilitate seamless collaboration between human and machine intelligence.
It includes three key steps:
\begin{enumerate}
    \item \textbf{Map}:
    As a user types, the system maps the current phrase to a list of potential corresponding code elements.

    \item \textbf{Assemble}:
    For each code element,
    the system assembles it with the existing code 
    by comparing the linguistic relationships between 
    the current phrase and previous phrases
    against a predefined rule set.
    Valid assemblies are presented as suggestions and previewed for the user in the code editor upon selection.

    \item \textbf{Preserve}:
    Once the user accepts a suggestion, 
    the system inserts the corresponding code elements to the code editor,
    and completes the typing phrase accordingly.
    The association between the phrase and the code element is preserved.
\end{enumerate}
The user can submit their prompt and the resultant code sketch to LLMs at any time to get the completed code.
We next explain the details of these three steps.




\subsection{Map}
When a user types a phrase, we aim to map it to potential corresponding code elements.
We identify that a phrase can refer to a code element in two distinct ways -- \emph{independently} or \emph{dependently} -- and thus employ two strategies to derive the potential mappings.

\paragraph{\textbf{Independent Phrase: Matching Predefined Code Snippets or Existing Code}}
A phrase can independently point to specific code elements using exact identifiers (e.g., variable or function names) or reserved keywords.
For example, in the prompt \emph{``The app should liste ...''}, the word \colorbox{mypink}{\texttt{app}} directly refers to a variable in the code.
Thus, we use fuzzy string matching to match the phrase with
two sources of code elements:

    \begin{itemize}
        \item \emph{Predefined Code Snippets.} 
        We provide a list of code snippets as potential matches for the typed word. 
        A code snippet is a small block of re-usable code, ranging from simple code fragments to complex function or class templates.
        Modern IDEs often provide predefined snippets.
        For example, \autoref{fig:ui_exp} shows a snippet for an API function from the Python Flask library~\cite{flask}.

        \item \emph{Existing Code.}
        We also dynamically parses the existing code to suggest identifiers, such as function names or variables, as potential matches for the typed word.       
    \end{itemize}

\paragraph{\textbf{Dependent Phrase: Inferring Based on Linguistic Features}}

A phrase can also refer to code elements based on its relationship with other phrases within the prompt.
This type of phrase requires linguistic context to resolve the code element it refers to. 
We use two NLP techniques to infer the corresponding code elements:

    \begin{itemize}
        \item \emph{Part-of-Speech.}
        In user's prompt, nouns usually indicate variables, while verbs suggest function calls.
        Recognizing these patterns, 
        we employ a \emph{part-of-speech tagging}~\cite{petrov2011universal} of the typed phrase and map it to a variable or function accordingly.
        For example, \autoref{fig:teaser}a shows that the system maps the words \colorbox{mypink}{\texttt{reads}} (a verb) and \colorbox{mypink}{\texttt{JSON}} (a noun) to a function call and an argument, respectively.
        
        \item \emph{Co-reference.}
        Certain phrases, such as \emph{``it''} or \emph{``the function''}, commonly refer back to other specific phrases mentioned earlier in the user's prompt. 
        Through \emph{co-reference analysis}~\cite{sukthanker2020anaphora}, we map these phrases to the corresponding code elements of the earlier 
        phrases they refer to.
    \end{itemize}

\subsection{Assemble}
\label{sec:assemble}

\begin{figure}[t]
    \centering
    \includegraphics[width=0.99\columnwidth]{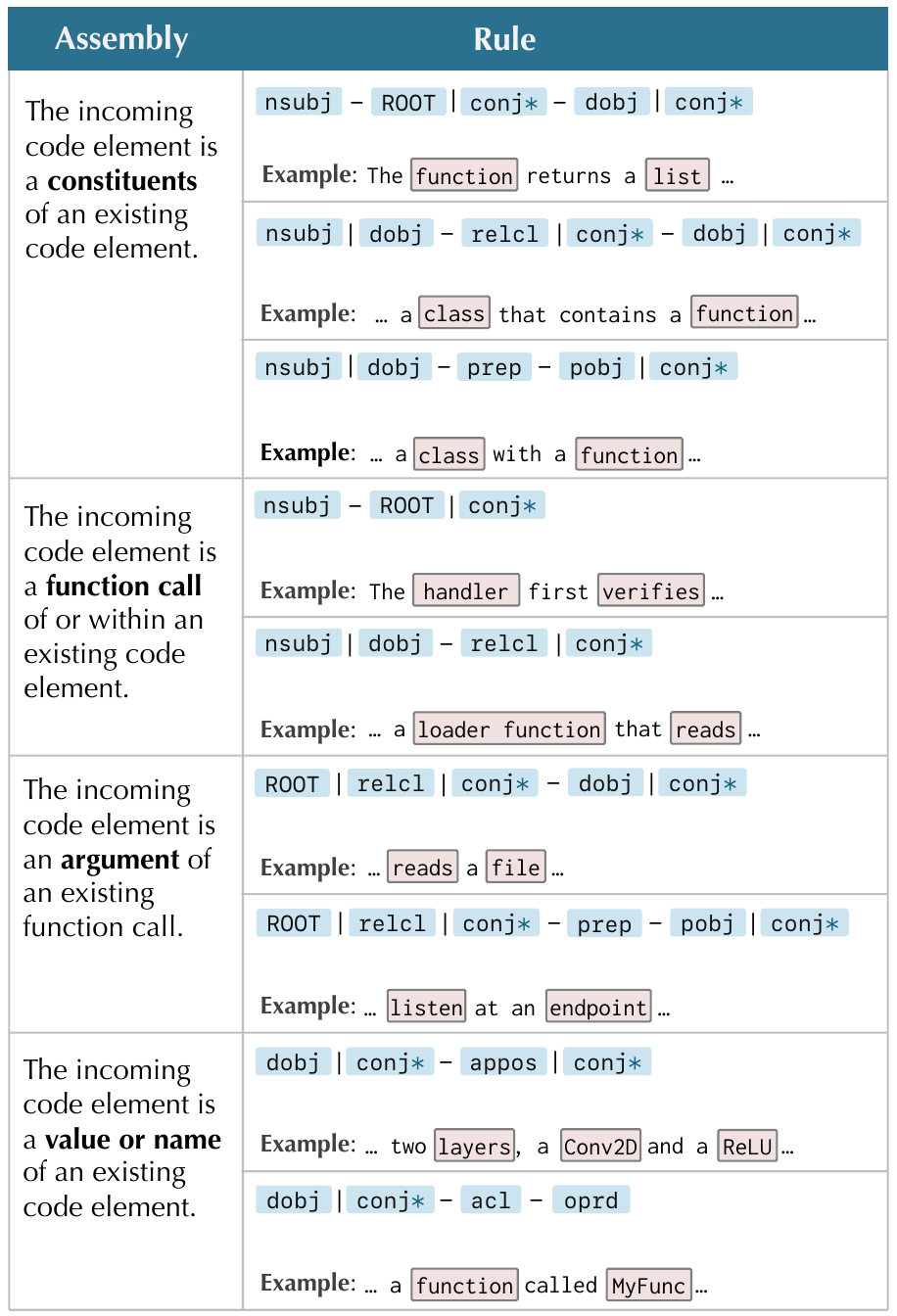}
    \caption{Example rules for code assembly, each represented as a regular expression that specifies the linguistic relationship between the incoming phrase and an existing code-associated phrase within the dependency tree. The labels used (e.g, \colorbox{myblue}{\texttt{nsubj}} and \colorbox{myblue}{\texttt{dobj}}) are established terms in NLP.}
    \label{fig:rules}
\end{figure}

Once a phrase is mapped to a potential code element, 
the next challenge is assembling this element with other existing code elements. 
To achieve this, we leverage the linguistic structures of the prompt that indicate relationships between code elements. 
Specifically, given the current phrase and its corresponding code elements,
our approach involves three main steps:
\begin{enumerate}
    \item \textbf{Dependency Parsing}.
    We begin with \emph{dependency parsing}~\cite{qi2019universal} to uncover the linguistic relationships between the current phrase and other phrases in the prompt. 
    This parsing technique organizes phrases into a dependency tree,
    showing how each phrase grammatically relates to and depends on its parent. The links above the words in \autoref{fig:teaser}a shows an example of dependency tree.

    \item \textbf{Rule Matching}.
    We then iteratively exam the linguistic relationships between the current phrase to other phrases associated with code elements following a reverse reading order. If a relationship aligns with a predefined rule in our rule set, we proceed with the assembly. 
    \autoref{fig:rules} shows a list of rule examples implemented in our current system.

    \item  \textbf{AST Assembly}.
    Instead of merely assembling plain text, we construct the Abstract Syntax Tree (AST) of these elements, 
    which provides a tree-like structure of the code that facilitates easier management.
    Each rule in our rule set also specifies how the ASTs of the two code elements involved in the rule should be constructed. 
    This process also adheres to the programming language's  grammar, 
    discarding any assemblies that do not comply.

\end{enumerate}
Finally, only the validated assemblies will be suggested to the user.
The user can select a suggestion (\autoref{fig:ui_exp}) and review how the code element will be inserted into the code editor.

\subsection{Preserve}
Once a user accepts a suggestion, 
our system preserves the established association between the phrase and its corresponding code element. 
This preserved association is crucial for the system to support incrementally building the code sketch. 
As the user inputs additional words, our system utilizes these prior associations to assemble the newly coming code elements as introduce in Sec.~\ref{sec:assemble}.

\subsection{Guiding LLMs in Code Generation}
The derived code sketch not only provides instant, incremental feedback to the user but also can guide the subsequent code generation process.
For example, we fill the prompt template below with the prompt and code sketch shown in \autoref{fig:teaser}b. 

\begin{coloredquotation}
    \textbf{Prompt template:} \\
    Based on the description and a code sketch (which is incomplete and buggy) below, do your best to complete the code with **minimal** editions. Generate the code within a triple quotes: \textasciigrave\textasciigrave\textasciigrave python ... \textasciigrave\textasciigrave\textasciigrave\\
    
    \#\#\# Code Description\\
    \{ \texttt{prompt} \} \\

    \#\#\# Current Code Draft\\
    \{ \texttt{code sketch} \} 
\end{coloredquotation}

\noindent
Once the LLM responds, we can streamingly display the edits made by the LLM on the code sketch, as illustrated in \autoref{fig:teaser}c. This process aids users in better understanding the generated code, as the sketch previews the code during the prompt-typing process.
\section{Preliminary User Feedback}

We implement our approach in a technology probe to gather preliminary feedback from colleagues who were unfamiliar with the project. 
A formal study is planned to further evaluate the usability of our approach, identify areas for improvement, and explore its real-world applications. 
Below, we briefly outline the implementation of the probe and the initial feedback received.

\paragraph{\textbf{Implementation.}}
We implemented our approach using a web-based code editor built with React~\cite{react}, CodeMirror~\cite{codemirror}, and TypeScript~\cite{typescript}, with NLP techniques powered by SpaCy~\cite{spacy}.
The ASTs for the code are obtained using Tree Sitter~\cite{treesitter}, an incremental parsing library. 
Furthermore, a Flask-based backend server was crafted to manage user prompts and facilitate interactions with the GPT-4 API from OpenAI. 
While currently supporting only Python, we envision extending our implementation to additional programming languages in subsequent iterations.

\paragraph{\textbf{User Feedback.}}
Feedback on the internal review was highly positive about the sketch, noting that it reduces the cognitive load for users by visualizing the potential structure of the code. 
Interestingly, users found the code sketch useful even when the assembly of code elements was not entirely accurate (e.g., a function inserted with incorrect indentation) because they expected the LLM to correct these errors. 
Additionally, users reported that it helps them better control the LLM. We believe this is due to two factors: 1) the code sketch provides an outline for the LLM, and 2) the feedback nudges users to write longer and more detailed prompts. 
The most significant concern regarding our approach is its robustness, which is unsurprising since we have only implemented a minimal set of rules in this technology probe. 
We anticipate that this issue can be alleviated with better engineering and more comprehensive rule sets. 
Feedback also indicated that sometimes the action of suggestion acceptance can be disruptive, particularly when the default suggestion is the intended one.
\section{Discussion and Future Work}
\vspace{-1mm}
\paragraph{\textbf{Target Contexts for Language-Oriented Code Sketching}}
Previous research~\cite{barke2023grounded} has found that the use of AI coding assistants varies according to the user's clarity about the final code solution. For example, Copilot Chat is favored for exploratory coding when the solution is unclear, whereas Copilot Code Completion is preferred for more targeted coding tasks with clearer objectives.
Our approach is particularly effective when users possess an intermediate level of clarity about their coding objectives. 
This typically occurs when users have a general understanding of the necessary functions, variables, or code structures, yet lack details on how to implement these elements. 
For example, 
a user may know they need to create a \emph{``REST API with a GET handler to read from a database''},
but are uncertain about the specifics of database interaction. 
In such cases, users frequently reference code elements as they refine their ideas and align the emerging code with their overall architectural goals. 

\begin{figure}[h]
    \centering
    \vspace{-2mm}
    \includegraphics[width=0.85\columnwidth]{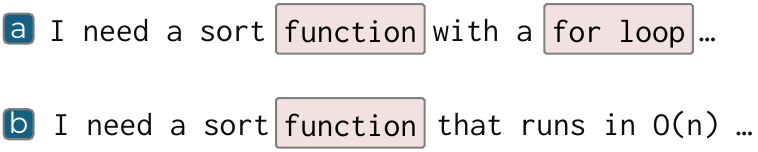}
    \caption{Our approach is more effective with prompts (a) that include more code-relevant phrases, allowing for the creation of more detailed code sketches, in contrast to prompts (b) that contain fewer such phrases.
    }
    \vspace{-2mm}
    \label{fig:ast_rate}
\end{figure}

\vspace{-1mm}
\paragraph{\textbf{Effectiveness Depends on Phrase-Code Correspondence}}
The effectiveness of our approach depends on the number of phrases that correspond to code elements. 
Intuitively, if a prompt contains more phrases that can be mapped to code elements, we can derive a more detailed code sketch.
For example, \autoref{fig:ast_rate}a contains two code-relevant phrases, while \autoref{fig:ast_rate}b contains only one. Consequently, our approach can create a more detailed sketch for the first prompt. 
Nonetheless, even in the worst case---where no sketch can be derived for a prompt---our approach will simply default to a conventional LLM-based text-to-code system.

\vspace{-1mm}
\paragraph{\textbf{Limitations.}}
Our approach is not without its limitations. 
As previously mentioned,
the effectiveness of our approach depends on the number of phrases that are correspondent to code elements within the user's prompt.
At present, our system supports only a minimal set of rules, which means that prompts that are ambiguous, unstructured, or overly generic may result in less precise code sketches. 
While it’s theoretically possible to expand our rule set in the future, the practicality of such enhancements is yet to be determined, and scaling up may reveal additional challenges.
Moreover, our current system is limited to English and Python only.

\vspace{-1mm}
\paragraph{\textbf{Future Work.}}
The transformation from a natural language utterance to a linguistic dependency tree, and ultimately to an AST, is essentially a parsing process.
We aim to streamline this by developing a compiler that would automate these steps and move beyond our current handcrafted, rule-based methodology, thereby increasing the system's robustness. 
By using code as an intermediate representation, the concept of \emph{sketch-then-generate} could be expanded to other domains, such as graphic design.
Furthermore, we envision that our approach could blur the distinctions between natural language and programming languages, opening up opportunities to apply features developed for programming languages to natural language prompts, such as static analysis, refactoring, and other enhancements. The latest updates will be posted here: \url{https://lang-sketch.github.io/}


\bibliographystyle{ACM-Reference-Format}
\bibliography{main.bib}

\end{document}